\documentclass[conference]{IEEEtran}
\IEEEoverridecommandlockouts
\usepackage{cite}
\usepackage{amsmath,amssymb,amsfonts}
\usepackage{algorithmic}
\usepackage{graphicx}
\usepackage{textcomp}
\usepackage{xcolor}
\usepackage{hyperref}
\usepackage{tikz}

\hypersetup{
    colorlinks=true,
    linkcolor=black,
    urlcolor=black,
    citecolor=black,
    pdfborder={0 0 0}
}

\definecolor{orcidlogocol}{HTML}{A6CE39}
\newcommand{\orcidlink}[1]{\href{https://orcid.org/#1}{\tikz[baseline=-0.5ex, x=1em, y=1em]\draw[orcidlogocol, fill=orcidlogocol] (0,0) circle (0.5) node[white] {\fontfamily{phv}\selectfont\fontsize{4}{4}\selectfont ID};}}

\def\BibTeX{{\rm B\kern-.05em{\sc i\kern-.025em b}\kern-.08em
    T\kern-.1667em\lower.7ex\hbox{E}\kern-.125emX}}
\begin{document}

\title{eFPE: Design, Implementation, and Evaluation of a Lightweight Format-Preserving Encryption Algorithm for Embedded Systems\\
}

\author{
\IEEEauthorblockN{Nishant Vasantkumar Hegde\textsuperscript{1,*}\orcidlink{0009-0009-8163-4489}}
\IEEEauthorblockA{\textit{Computer Science and Engineering} \\
\textit{RV College of Engineering}\\
Bengaluru, INDIA \\
nishantvh.cy23@rvce.edu.in}
\and
\IEEEauthorblockN{Suneesh Bare}
\IEEEauthorblockA{\textit{Computer Science and Engineering} \\
\textit{RV College of Engineering}\\
Bengaluru, INDIA \\
suneeshbare.cy23@rvce.edu.in}
\and
\IEEEauthorblockN{K B Ramesh}
\IEEEauthorblockA{\textit{Electronics and Communication Engineering} \\
\textit{RV College of Engineering}\\
Bengaluru, INDIA \\
rameshkb@rvce.edu.in}
\and
\IEEEauthorblockN{Aamir Ibrahim}
\IEEEauthorblockA{\textit{Computer Science and Engineering} \\
\textit{RV College of Engineering}\\
Bengaluru, INDIA \\
aamiribrahim.cy23@rvce.edu.in}
}

\maketitle

\begingroup
\renewcommand\thefootnote{}
\footnotetext{
\textsuperscript{1}First author \\
\textsuperscript{*}Corresponding author. Email: hegde.nishant2005@gmail.com
}
\endgroup

\begin{abstract}
Resource-constrained embedded systems demand secure yet lightweight data protection, particularly when data formats must be preserved. This paper introduces eFPE (Enhanced Format-Preserving Encryption), an 8-round Feistel cipher featuring a "novel lightweight Pseudorandom Function (PRF)" specifically designed for this domain. The PRF, architected with an efficient two-iteration structure of AES-inspired operations (byte-substitution, keyed XOR, and byte-rotation), underpins eFPE's ability to directly encrypt even-length decimal strings without padding or complex conversions, while aiming for IND-CCA2 security under standard assumptions. Implemented and evaluated on an ARM7TDMI LPC2148 microcontroller using Keil µVision 4, eFPE demonstrates the efficacy of its targeted design: a total firmware Read-Only Memory (ROM) footprint of 4.73~kB and Random Access Memory (RAM) usage of 1.34~kB. The core eFPE algorithm module itself is notably compact, requiring only 3.55~kB ROM and 116~B RAM. These characteristics make eFPE a distinct and highly suitable solution for applications like financial terminals, medical sensors, and industrial IoT devices where data format integrity, minimal resource footprint, and low operational latency are paramount.
\end{abstract}


\section{Introduction}
\label{sec:introduction}
The proliferation of Internet of Things (IoT) devices and connected systems presents significant challenges in securing sensitive data, particularly on platforms with constrained computational power and memory \cite{Thakor2021Lightweight, Radhakrishnan2024Efficiency}. Traditional cryptographic algorithms, while robust, often prove too resource-intensive for such embedded environments. Format-Preserving Encryption (FPE) \cite{Bellare2009Format, Lampe2012Asymptotically} offers a compelling solution by enabling the encryption of data (e.g., credit card numbers, patient IDs) while maintaining its original structure and length. This property is crucial for ensuring compatibility with legacy systems and existing data processing pipelines. However, even standardized FPE methods like FF1 and FF3, as stipulated by NIST \cite{Dworkin2016NISTSP80038G}, often rely on full implementations of block ciphers like AES, which can be computationally demanding for many microcontrollers.

To address these limitations, this paper introduces eFPE (Enhanced Format-Preserving Encryption), a lightweight cipher whose primary novelty lies in its "specifically architected cryptographic core designed for resource-constrained embedded FPE applications." eFPE employs an 8-round balanced Feistel network driven by a "novel, lightweight AES-inspired Pseudorandom Function (PRF)." This PRF utilizes an efficient two-iteration structure of carefully selected byte-level operations (byte-substitution via the AES S-box, keyed XOR, and byte-rotation) to achieve a strong balance between security and minimal computational overhead. eFPE is optimized for the direct encryption of any even-length decimal string without requiring padding or complex format conversions. The primary design goals are to minimize memory footprint (ROM and RAM), reduce computational complexity for low-latency operation, and ensure robust security guarantees (aiming for IND-CCA2), all while strictly preserving the data's numeric format.

Our contributions include:
\begin{enumerate}
    \item The design and specification of the eFPE algorithm, prominently featuring its "novel lightweight PRF construction" and its integration into an 8-round Feistel network tailored for efficient decimal FPE. This includes the resource-efficient key schedule derived using the same PRF logic.
    \item A comprehensive performance evaluation of the eFPE implementation on an NXP LPC2148 microcontroller, quantifying its memory usage (ROM/RAM) and demonstrating the practical benefits of its lightweight design.
    \item The validation of a complete eFPE-based system prototype, showcasing its applicability and operational characteristics in a realistic embedded context.
\end{enumerate}

The results confirm eFPE's suitability as a distinct solution for securing numeric data in applications such as financial payment terminals, medical sensor devices, and industrial IoT nodes, where efficiency, format integrity, and low resource consumption are paramount.

The remainder of this paper is organized as follows: Section~\ref{sec:algorithm_design} details the eFPE algorithm. Section~\ref{sec:security_analysis} provides its security analysis. Section~\ref{sec:implementation} describes the implementation and experimental setup used for validation. Section~\ref{sec:performance_evaluation} presents the performance evaluation of eFPE. Section~\ref{sec:comparative_analysis} offers a qualitative comparative analysis. Finally, Section~\ref{sec:conclusion} concludes the paper and outlines future work.

\section{eFPE Algorithm Design}
\label{sec:algorithm_design}
\begin{figure}[t]
\centering
\includegraphics[width=\columnwidth]{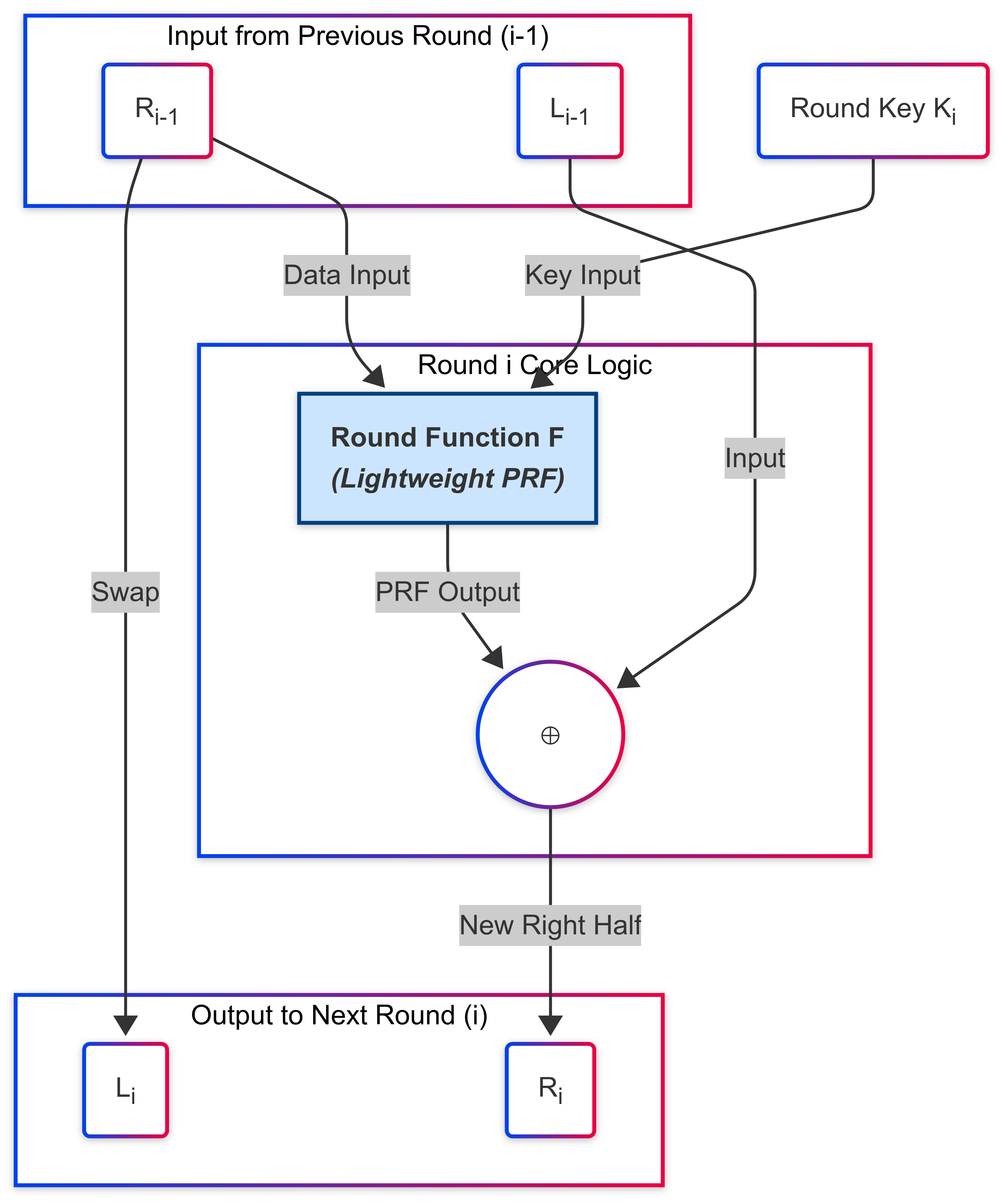} 
\caption{The data flow of a single round ($i$) in the eFPE Feistel network. The right half ($R_{i-1}$) and the round key ($K_i$) are processed by the lightweight PRF (F). The result is XORed with the left half ($L_{i-1}$) to produce the new right half ($R_i$), while the original right half is swapped to become the new left half ($L_i$).}
\label{fig:feistel_round}
\end{figure}

The eFPE algorithm is architected as an N-round (specifically N=8 for our implementation) balanced Feistel network, designed to directly encrypt even-length decimal strings without padding or format conversion. Central to Feistel ciphers \cite{Feistel1973Cryptography}, an n-digit input decimal string is first divided into two equal m-digit halves, L (left) and R (right), where m = n/2. The data flow of this transformation for a single round $i$ is illustrated in Fig.~\ref{fig:feistel_round}. Each round applies the following transformation:
\begin{align}
    T_i &= L_{i-1} \nonumber \\
    L_i &= R_{i-1} \\
    R_i &= T_i \oplus F(R_{i-1}, K_i, i) \pmod{10^m} \nonumber
    \label{eq:feistel_update}
\end{align}
In this transformation, $L_{i-1}$ and $R_{i-1}$ are the left and right halves from the previous round. $T_i$ is a temporary variable holding the value of $L_{i-1}$ before it is overwritten. $K_i$ is the unique key for round $i$, and $F$ is the round function. The modulo $10^m$ operation is critical for ensuring that the resulting right half, $R_i$, retains its m-digit numeric property, thereby preserving the overall data format.

\subsection{Round Function (F)}
The security and efficiency of eFPE heavily rely on its round function $F$. It is defined as:
\begin{equation}
F(R, K_i, i) = \text{PRF}_{K_i} (R \,||\, i) \pmod{10^m}
\label{eq:round_function}
\end{equation}
Here, the right half-block $R$ from the input is concatenated ($||$) with the current round index $i$. This combined value serves as the input to a keyed Pseudorandom Function (PRF). The PRF is keyed with the round key $K_i$, and its output is subsequently reduced modulo $10^m$ to match the domain size of the half-block. This PRF is inspired by AES operations but is significantly simplified for resource-constrained devices. It involves two iterations of the following three steps applied to an internal 32-bit state:
\begin{enumerate}
    \item \textbf{SubBytes:} A non-linear byte substitution using the standard 8x8 AES S-box. This step is crucial for providing confusion and resistance against differential and linear cryptanalysis.
    \item \textbf{AddRoundKey:} A bitwise XOR operation between the internal state and a round-specific subkey.
    \item \textbf{Rotate:} A cyclic left rotation of the bytes within the 32-bit state. This promotes diffusion by spreading the influence of input bits across multiple output bits.
\end{enumerate}

\subsection{Key Schedule}
eFPE employs a lightweight key schedule to derive the N unique round keys $K_i$ from a single 128-bit master key (MK). Each round key is generated by an application of the same PRF structure used in the round function, thus reusing code to minimize the memory footprint:
\begin{equation}
K_i = \text{PRF}_{\text{MK}} (i \,||\, \text{const})
\label{eq:key_schedule}
\end{equation}
In this equation, $K_i$ is the generated round key for round $i$. The PRF is keyed with the master key, $\text{MK}$. Its input is the concatenation ($||$) of the round index $i$ and a predefined constant value (`const`, e.g., 0xA5A5A5A5). Using the unique round index $i$ as input ensures that each generated key $K_i$ is distinct, which is vital for resisting attacks like slide attacks.

\subsection{Encryption and Decryption Process Summary}
For encryption, the input plaintext decimal string is converted into two numeric values, L and R. These then pass through the N Feistel rounds as described. After the final round, the resulting L and R values are converted back into their decimal string representations and concatenated to form the ciphertext. Decryption follows the exact same process but with the Feistel rounds applied in reverse order (from N-1 down to 0), using the same sequence of round keys. The inherent reversibility of the Feistel structure guarantees that the original plaintext is perfectly recovered.

\section{Security Analysis}
\label{sec:security_analysis}
The security of eFPE is founded on the established properties of Feistel networks and the cryptographic strength of its custom lightweight Pseudorandom Function (PRF). The overall security is achieved through a multi-layered design: the non-linearity and diffusion from the PRF's AES-inspired operations provide confusion; the 8-round Feistel structure offers formal security guarantees; and a key schedule that uses round indices prevents round-to-round vulnerabilities. This composite approach provides robust security suitable for resource-constrained environments while adhering to strict resource constraints.

\subsection{Theoretical Security Model}
eFPE is analyzed under the standard notion of Indistinguishability under Chosen-Ciphertext Attack (IND-CCA2). For a Feistel network, security largely depends on the pseudorandomness of its round function. Assuming the underlying PRF used in eFPE's round function is secure (i.e., computationally indistinguishable from a truly random function), the security of an N-round Feistel cipher against an adversary $A$ making $q$ queries can be bounded. A common result, such as Luby-Rackoff, shows that for a sufficient number of rounds (e.g., 3-4 with ideal random functions, more with PRFs), the Feistel construction can achieve security close to that of a random permutation. The security bound for eFPE, considering $N=8$ rounds, is primarily influenced by the PRF's security advantage ($\text{Adv}_{\text{PRF}}$) and the number of queries relative to the block size: $\text{Adv}_{\text{eFPE}}^{\text{IND-CCA2}}(A) \leq q^2 / 2^k + \text{Adv}_{\text{PRF}}(B)$.

To clarify the parameters in this security bound:
\begin{itemize}
    \item \textbf{$\text{Adv}_{\text{eFPE}}^{\text{IND-CCA2}}(A)$}: Represents the advantage of an adversary $A$ in successfully breaking the IND-CCA2 security of the eFPE cipher. This measures the probability of the adversary winning the security game beyond random guessing.
    \item \textbf{$q$}: The total number of queries (e.g., requests for encryption or decryption) that the adversary is permitted to make to the system.
    \item \textbf{$k$}: The bit-length of a half-block, which is derived from its $m$ decimal digits ($k \approx m \times \log_2(10)$). The $q^2 / 2^k$ term represents the security loss due to the birthday bound on the block size.
    \item \textbf{$\text{Adv}_{\text{PRF}}(B)$}: Denotes the advantage of a simulator or distinguisher $B$ in distinguishing the output of the eFPE PRF from that of a truly random function.
    \item \textbf{$B$}: The simulator for the PRF, an algorithmic entity used in the security proof to interact with the PRF distinguisher.
\end{itemize}

The 8 rounds are chosen to provide a good balance between achieving sufficient diffusion and confusion for security, and maintaining computational efficiency.

\subsection{Resistance to Common Cryptanalytic Attacks}
eFPE incorporates design features to mitigate several well-known cryptanalytic attacks. To defend against \textbf{Differential and Linear Cryptanalysis}, which exploits statistical non-uniformities in a cipher's operations, eFPE relies on the strong non-linearity of the chosen AES S-box. This property, combined with the diffusion provided by the Rotate step over 8 rounds, ensures that any input differences or linear approximations rapidly dissipate, rendering the search for exploitable characteristics computationally infeasible. \textbf{Slide Attacks}, which exploit self-similarity in ciphers where the same round function and key are used repeatedly, are mitigated by ensuring each round is distinct. This is achieved by incorporating the round index $i$ as an input to both the round function $F$ and the key schedule that generates unique round keys $K_i$. Finally, the complexity of \textbf{Meet-in-the-Middle Attacks}, which attempt to find an intermediate state by matching encryption from one end with decryption from the other, is significantly increased beyond practical limits by the use of $N=8$ rounds and distinct, cryptographically derived keys, making it difficult to isolate intermediate states.

\subsection{Security versus Block Size}
The security of eFPE, like other FPE schemes operating on a fixed radix, is inherently tied to the size of its operational domain. For a balanced Feistel network processing an $n$-digit decimal input, the input is split into two $m$-digit halves, where $m = n/2$. The strength against certain attacks, particularly brute-force attacks against the possible values of a half-block or attacks related to the birthday bound, is determined by the number of possible values a half-block can take, which is $10^m$.

The effective security level in bits for such a half-block can be estimated as $\log_2(10^m) = m \times \log_2(10)$. Since $\log_2(10) \approx 3.3219$, the effective security is approximately $m \times 3.3219$ bits. The corresponding attack complexity to guess a half-block value would be on the order of $10^m$ operations (or $2^{m \times \log_2(10)}$ operations).

\begin{table}[h!]
\centering
\caption{EFFECTIVE SECURITY LEVELS FOR eFPE (FULL eFPE BLOCK IS 2m DIGITS)}
\label{tab:security_levels}
\begin{tabular}{ccc}
\hline
\textit{\textbf{Half-Block Size (m)}} & \textit{\textbf{Effective Security}} & \textit{\textbf{Attack Complexity}} \\
\textit{\textbf{(digits)}}             & \textit{\textbf{(bits)}}             & \textit{\textbf{(approx. \boldmath$10^m$ ops)}}      \\
\hline
4                                     & $\sim$13.3                           & $\sim 10^{4}$                        \\
8                                     & $\sim$26.6                           & $\sim 10^{8}$                        \\
12                                    & $\sim$39.9                           & $\sim 10^{12}$                       \\
16                                    & $\sim$53.2                           & $\sim 10^{16}$                       \\
\end{tabular}
\end{table}

Table~\ref{tab:security_levels} presents these effective security levels and approximate attack complexities for different half-block sizes ($m$) used within eFPE. It is important to note that the "Half-Block Size (m) (digits)" in Table~\ref{tab:security_levels} refers to this half-block size. The full plaintext block size would be $2m$. eFPE's design allows for configurable block sizes, enabling implementers to choose a security level appropriate for their application's risk profile and performance requirements. For instance, using $m=8$ (a 16-digit full block) provides significantly more security than $m=2$ (a 4-digit full block, often used for PINs).

\section{Implementation and Experimental Setup}
\label{sec:implementation}
\begin{figure}[t]
\centering
\includegraphics[width=0.95\columnwidth]{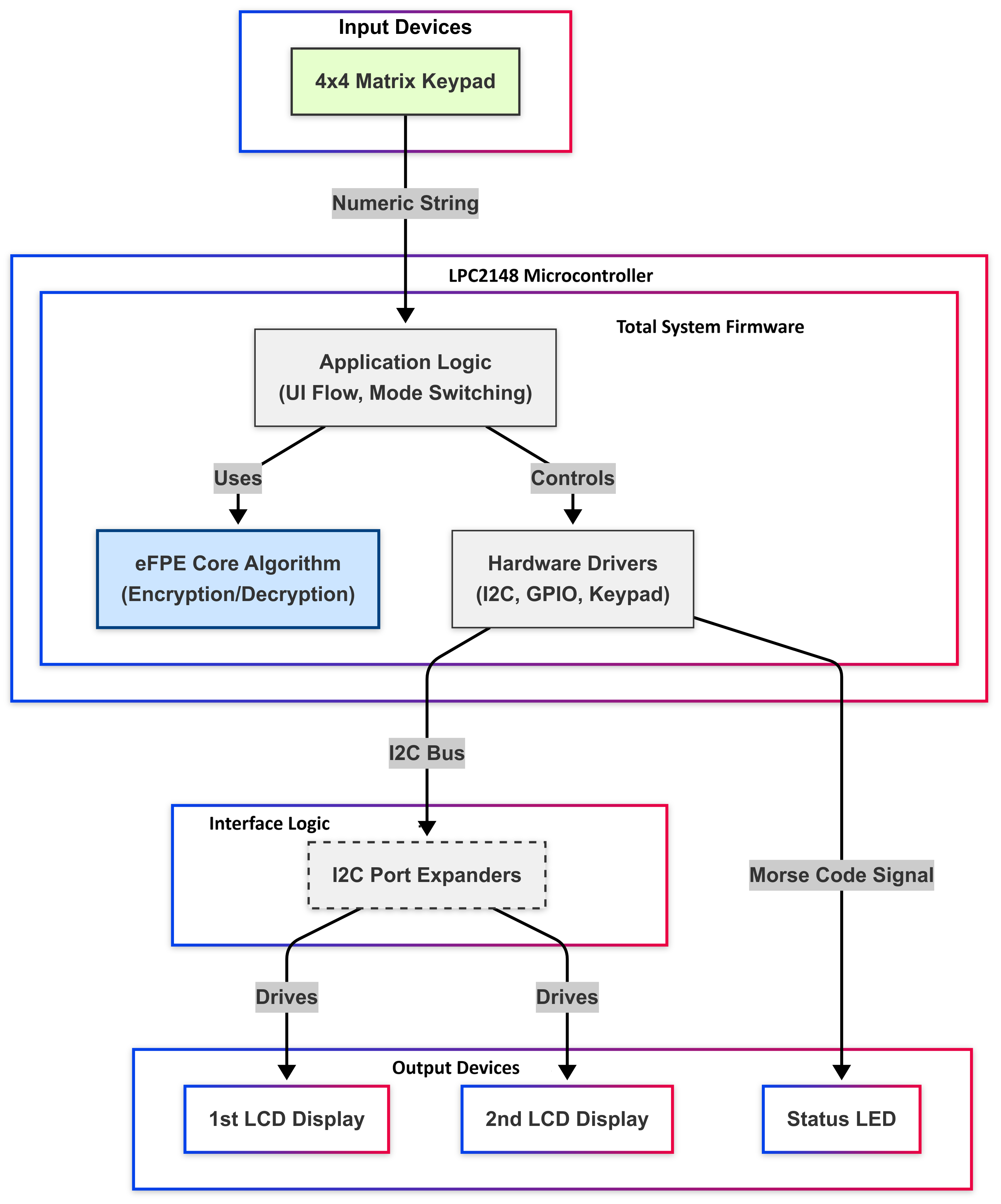} 
\caption{System architecture of the eFPE prototype. The diagram shows the hardware components and the internal software structure of the firmware running on the LPC2148 MCU. The 'Total System Firmware' comprises the Application Logic, Hardware Drivers, and the core cryptographic module, providing crucial context for the performance analysis in Section~V.}
\label{fig:system_architecture}
\end{figure}

To evaluate the performance of eFPE and validate its practical applicability, the algorithm was implemented on a representative microcontroller and integrated into a complete system prototype. The overall system architecture, detailing the interaction between the internal software components and the external hardware peripherals, is depicted in Fig.~\ref{fig:system_architecture}. This architecture forms the basis for the validation and performance characterization described below.

\subsection{Hardware and Development Environment}
The core eFPE algorithm was implemented in C, targeting the NXP LPC2148, a common microcontroller featuring an ARM7TDMI-S core. This platform was chosen as it represents a realistic target for resource-constrained embedded applications. Firmware development was conducted using the Keil µVision 4 IDE with the ARM RealView toolchain. The firmware was designed with a modular organization (e.g., main control logic, keypad scanning, I$^2$C peripheral drivers, eFPE core) to facilitate unit testing and maintainability.

\subsection{Validation and Evaluation Methodology}
A two-stage process was employed to ensure both functional correctness and to accurately characterize performance:
\begin{enumerate}
    \item \textbf{Functional Validation through Simulation:} The entire system's hardware integration and real-time behavior were first validated in the Proteus Design Suite 8. This co-simulation environment allowed for comprehensive system-level testing of the firmware's logic, including input handling and peripheral communication, before deployment on physical hardware.
    \item \textbf{Performance Characterization:} Following successful functional validation, the evaluation focused on quantifying the memory footprint. The complete, validated firmware was compiled using the Keil toolchain. The precise memory usage figures reported in Section~V were extracted directly from the linker-generated map files, providing an accurate and verifiable characterization of the algorithm's resource consumption.
\end{enumerate}

\subsection{System Prototype Operation}
The validated prototype operates in two distinct modes: Encrypt (ENC) and Decrypt (DEC), toggled by a specific key. In ENC mode, as a user enters an even-length decimal string, it is processed by eFPE, and the resulting ciphertext is displayed. In DEC mode, entered ciphertext is processed similarly to recover the plaintext. For both modes, an LED provides alternative visual feedback via Morse code \cite{Galamgam2022MorseCode}.

\section{Performance Evaluation of eFPE}
\label{sec:performance_evaluation}
The performance characteristics of the developed eFPE implementation were meticulously analyzed based on its compilation for the NXP LPC2148 microcontroller, as detailed in Section~\ref{sec:implementation}. This evaluation focuses on eFPE's memory footprint, a critical factor for resource-constrained embedded systems, and discusses the implications for operational and energy efficiency.

\subsection{Memory Footprint Analysis}
The memory footprint of the complete eFPE system firmware was determined from the linker-generated map files. Table~\ref{tab:efpe_total_memory_footprint} summarizes the overall memory usage for the entire executable programmed onto the LPC2148. The total Read-Only Memory (ROM) usage was found to be 4728 bytes (4.73~kB), while the total Random Access Memory (RAM) usage was 1376 bytes (1.34~kB).

\begin{table}[h!]
\centering
\caption{EFPE TOTAL SYSTEM MEMORY FOOTPRINT ON LPC2148}
\label{tab:efpe_total_memory_footprint}
\begin{tabular}{lr}
\hline
\textit{\textbf{Memory Category}} & \textit{\textbf{Size (Bytes)}} \\[1.5pt]
\hline
Total Read-Only Memory (ROM)    & 4728 \\[1.5pt]
\quad (Code + Read-Only Data)   &      \\[1.5pt]
Total Read-Write Memory (RAM)   & 1376 \\[1.5pt]
\quad (RW Data + Zero-Init Data) &      \\
\end{tabular}
\end{table}

To provide further insight into the efficiency of the core cryptographic logic, Table~\ref{tab:efpe_core_module_footprint} details the memory contribution specifically from the \texttt{efpe\_code.o} object file. The eFPE core module itself contributes only 3552 bytes to ROM and a minimal 116 bytes to RAM.

\begin{table}[h!]
\centering
\caption{EFPE CORE ALGORITHM MODULE FOOTPRINT (\texttt{efpe\_code.o})}
\label{tab:efpe_core_module_footprint}
\begin{tabular}{lr}
\hline
\textit{\textbf{Component (\texttt{efpe\_code.o})}} & \textit{\textbf{Size (Bytes)}} \\[1.5pt]
\hline
Code (Executable instructions)      & 3176 \\[1.5pt]
Read-Only Data (Constants)          & 376  \\[1.5pt]
\textbf{Subtotal ROM (eFPE Core)}   & \textbf{3552} \\[1.5pt]
\hline
Read-Write Data (Init. Vars)      & 84 \\[1.5pt]
Zero-Initialized Data (Uninit. Vars) & 32 \\[1.5pt]
\textbf{Subtotal RAM (eFPE Core)}   & \textbf{116} \\
\end{tabular}
\end{table}

\subsection{Implied Operational Efficiency and Energy Considerations}
The compact memory footprint detailed above is a strong indicator of high operational efficiency. A smaller code size translates to a lower instruction count, which implies faster execution and reduced CPU utilization. While direct cycle counting was not the focus of this evaluation, this inherent leanness logically leads to more favorable energy consumption characteristics, a critical advantage for battery-powered and energy-sensitive embedded applications.

\subsection{Discussion of eFPE Resource Efficiency}
\begin{figure}[t]
\centering
\includegraphics[width=\columnwidth]{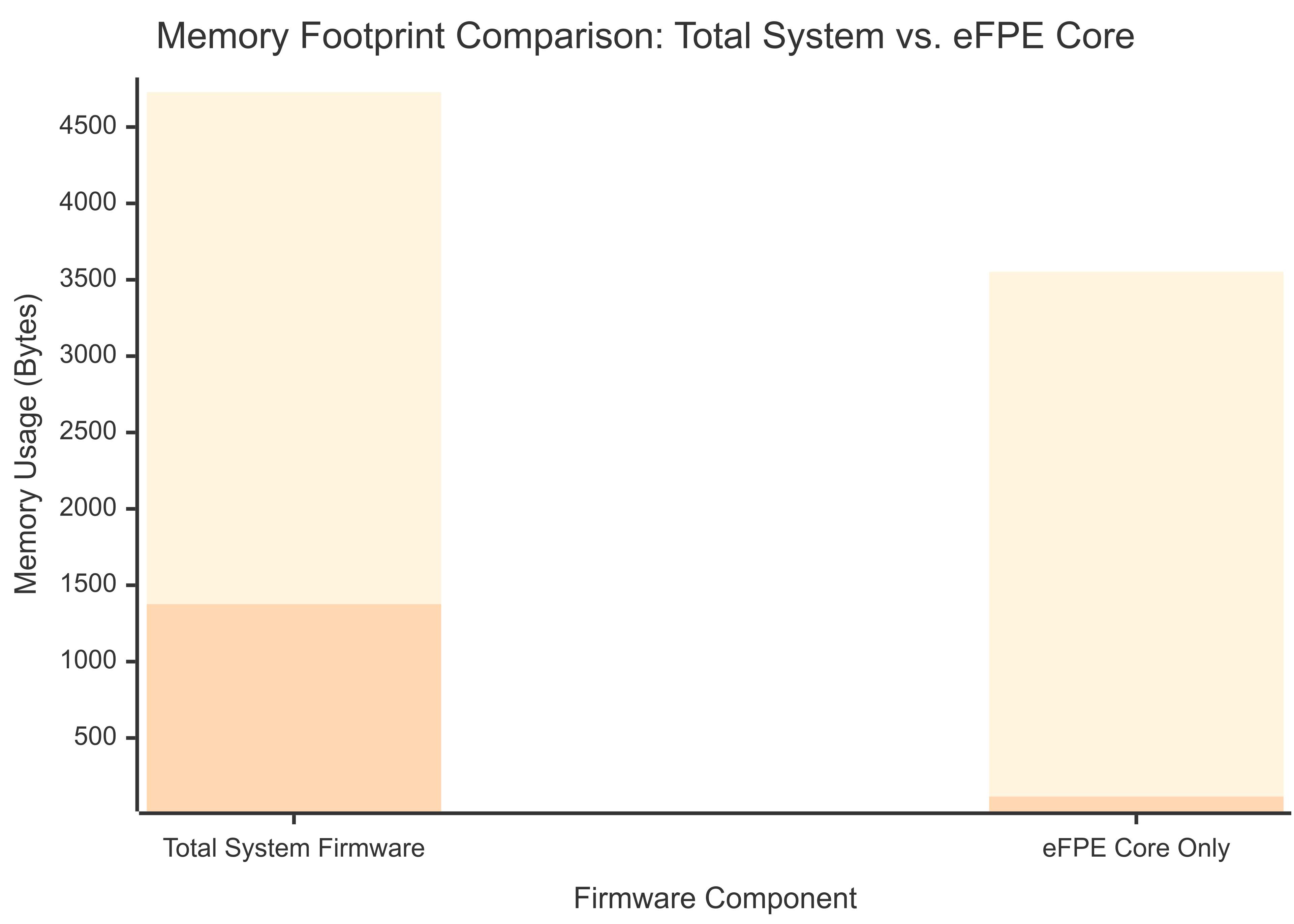} 
\caption{Memory footprint comparison of the Total System Firmware versus the eFPE Core only. The stacked bars show the breakdown of RAM (darker orange/lower segment) and ROM (lighter orange/upper segment). The chart visually emphasizes the minimal resource consumption of the core algorithm relative to the full system's overhead, particularly the very small RAM requirement.}
\label{fig:performance_chart}
\end{figure}

The memory analysis confirms eFPE's suitability for deployment on microcontrollers with stringent memory limitations. The total system footprint (under 5~kB ROM and 1.4~kB RAM) is well within typical capacities, and as shown in Fig.~\ref{fig:performance_chart}, the core algorithm module is exceptionally small. Requiring just 3.55~kB of ROM and 116 bytes of RAM, its compactness is a direct result of design choices like reusing the PRF logic. This resource efficiency is crucial for enabling secure FPE in diverse embedded scenarios without overburdening the system.

\section{Comparative Analysis}
\label{sec:comparative_analysis}

While this paper's primary focus is the empirical performance of the eFPE implementation (Section~\ref{sec:performance_evaluation}), it is useful to position its design philosophy relative to other cryptographic approaches. This section provides a conceptual comparison against two relevant categories: NIST-standardized FPE schemes and general-purpose Lightweight Block Ciphers (LWC).

\subsection{Context with NIST FPE Standards}
NIST FPE standards like FF1 and FF3 \cite{Dworkin2016NISTSP80038G} are robust and versatile, often built upon established ciphers like AES. They provide strong security for a wide variety of data types. However, their generality can lead to implementations that are resource-intensive for highly constrained microcontrollers. Table~\ref{tab:conceptual_nist_fpe} outlines a conceptual comparison between eFPE and the general characteristics of such standards. eFPE, by specializing in numeric data and employing a significantly lighter AES-inspired PRF, is architected to achieve a more optimized balance of features for its specific domain.

\begin{table}[h!]
\centering
\caption{CONCEPTUAL COMPARISON: EFPE VS. NIST FPE STANDARDS}
\label{tab:conceptual_nist_fpe}
\begin{tabular}{p{2cm} p{2.7cm} p{2.7cm}}
\hline
\textit{\textbf{Aspect}} & \textit{\textbf{NIST FPE (e.g., FF1)}} & \textit{\textbf{eFPE (Proposed)}} \\[1.5pt]
\hline
Primary Design Goal & General FPE, high security & Lightweight FPE, constrained devices \\[3pt]
Target Data Types & Alphanumeric, various charsets & Numeric (decimal strings) \\[3pt]
Underlying Crypto. Primitive & Full AES / std. block cipher & Lightweight AES-inspired PRF \\[3pt]
Resource Footprint (ROM/RAM) & Moderate to High & Low to Very Low \\[3pt]
Computational Complexity & Moderate to High & Low \\[3pt]
Integration Effort (Basic Numeric) & Standardized, larger library & Minimal, self-contained \\
\end{tabular}
\end{table}

\subsection{Context with Lightweight Block Ciphers (LWC)}
Lightweight block ciphers (LWC) such as PRESENT \cite{Bogdanov2007PRESENT} and SIMON \cite{Rashidi2020FlexibleSIMON} are highly optimized for minimal footprint and efficient binary data encryption. While their core logic is extremely compact, they do not inherently preserve data formats like decimal strings. As shown in Table~\ref{tab:conceptual_lwc}, using LWCs for FPE tasks typically requires additional layers for encoding, decoding, and padding, which adds complexity and resource overhead. In contrast, eFPE’s design directly incorporates format preservation for numeric strings, offering a more integrated and efficient solution for this specific task.

\begin{table}[h!]
\centering
\caption{CONCEPTUAL COMPARISON: EFPE VS. TYPICAL LWC}
\label{tab:conceptual_lwc}
\begin{tabular}{p{2.1cm} p{2.6cm} p{2.7cm}}
\hline
\textit{\textbf{Aspect}} & \textit{\textbf{Typical LWC (e.g., PRESENT, SIMON)}} & \textit{\textbf{eFPE (Proposed)}} \\[1.5pt]
\hline
Primary Design Goal & Min. footprint/energy for binary blocks & Lightweight FPE for numeric strings \\[3pt]
Native Data Type & Binary blocks & Numeric (decimal strings) \\[3pt]
Inherent Format Preservation & No (binary only) & Yes (decimal native) \\[3pt]
Need External Layers for FPE & Yes (encode/decode, padding) & No (FPE is native) \\[3pt]
Overall Complexity for FPE Task & Higher (core + format layers) & Lower (integrated) \\
\end{tabular}
\end{table}

In essence, eFPE occupies a niche where direct format preservation of numeric data is required on severely resource-constrained devices, offering a tailored approach that prioritizes low footprint and efficient processing for this specific task.

\section{Conclusion and Future Work}
\label{sec:conclusion}

This paper introduced eFPE (Enhanced Format-Preserving Encryption), a novel lightweight cryptographic scheme for securing numeric data in resource-constrained embedded systems. The core innovation is an efficiently architected 8-round Feistel network driven by a custom, lightweight AES-inspired Pseudorandom Function (PRF). The empirical evaluation on an ARM7TDMI microcontroller demonstrated the practical efficacy of this targeted design, achieving a total firmware footprint of 4.73~kB ROM and 1.34~kB RAM, with the core algorithm itself requiring an exceptionally small 3.55~kB ROM and 116~B RAM. These characteristics, supported by a validated system prototype, underscore eFPE's suitability for applications like financial terminals and IoT sensors where memory, latency, and format integrity are critical.

Future work will focus on extending eFPE's capabilities and enhancing its evaluation. This includes adding support for alphanumeric data, exploring hardware acceleration, and performing precise on-target cycle count and power profiling to quantitatively confirm its efficiency. Further investigation will address advanced system-level hardening. This involves exploring dynamic voltage and frequency scaling (DVFS) for energy reduction, architecting atomic state transitions to ensure data integrity against power failure, and implementing robust input validation and anti-tamper techniques on interfaces to mitigate timing and fault-injection attacks. Finally, continued security analysis against emerging cryptanalytic techniques remains essential for ensuring the long-term robustness of eFPE.

\section*{Acknowledgment} 
The authors express their gratitude to the Management of Rashtreeya Sikshana Samithi Trust (RSST), and the Principal and Vice Principal of RV College of Engineering, Bengaluru, India, for their continuous support and encouragement.


\end{document}